\documentclass[12pt]{article}
\pdfoutput=1

\usepackage[margin=3.5cm]{geometry}
\usepackage{amsmath,amssymb,graphicx}
\usepackage[dvipsnames]{xcolor}
\usepackage{hyperref}
\usepackage{multicol}
\usepackage{soul}

\renewcommand{\leq}{\leqslant}
\renewcommand{\geq}{\geqslant}

\numberwithin{equation}{section}

\begin{document}


\thispagestyle{empty}



\begin{center}
	
{\Large \bf Worldline formalism for a confined scalar field}\vspace{1cm}
	
{\large Olindo Corradini$^{1,2}$, James P.\ Edwards$^{3}$, Idrish Huet$^{4}$,\\[1mm] Lucas Manzo$^{5}$, and Pablo Pisani$^{6}$}\\\vspace{1cm}
$^{1}$\textit{Dipartimento di Scienze Fisiche, Informatiche e Matematiche,\\ Universit\`a degli Studi di Modena e Reggio Emilia,\\ Via Campi 213/A, I-41125 Modena, Italy}\\[2mm]
$^{2}$\textit{Istituto Nazionale di Fisica Nucleare, \\ Via Irnerio 46, I-40126 Bologna, Italy}\\[2mm]
$^{3}$\textit{\mbox{Instituto de F\'isica y Matem\'aticas, Universidad Michoacana de San Nicol\'as de Hidalgo,} 
 Edificio C-$3$, Apdo.\ Postal $2$-$82$, C.P.\ 58040, Morelia, Michoac\'an, M\'exico}\\[2mm]
$^{4}$\textit{Facultad de Ciencias en F\'isica y Matem\'aticas,\\Universidad Aut\'onoma de Chiapas,\\ Ciudad Universitaria, Tuxtla Guti\'errez 29050, M\'exico.}\\[2mm]
$^{5}$\textit{Departamento de F\'isica de la Facultad de Ciencias Exactas,\\Universidad Nacional de La Plata, CC 67 (1900) La Plata, Argentina.}\\[2mm] 
$^{6}$\textit{Instituto de F\'isica La Plata, CONICET and Universidad Nacional de La Plata, CC 67 (1900) La Plata, Argentina.}\\\vspace{3mm}
\texttt{olindo.corradini@unimore.it, jedwards@ifm.umich.mx, idrish@ifm.umich.mx, lucasmanzo@fisica.unlp.edu.ar, pisani@fisica.unlp.edu.ar}\vspace{7mm}

\end{center}

\begin{abstract}
\noindent The worldline formalism is a useful scheme in quantum field theory which has also become a powerful tool for numerical computations. The key ingredient in this formalism is the first quantization of an auxiliary point-particle whose transition amplitudes correspond to the heat-kernel of the operator of quantum fluctuations of the field theory. However, to study a quantum field which is confined within some boundaries one needs to restrict the path integration domain of the auxiliary point-particle to a specific subset of worldlines enclosed by those boundaries. We show how to implement this restriction for the case of a scalar field confined to the $D$-dimensional ball under Dirichlet and Neumann boundary conditions, and compute the first few heat-kernel coefficients as a verification of our construction. We argue that this approach could admit different generalizations.
\end{abstract}




\section{Introduction}

Since the influential works of Z.\ Bern and D.\ A.\ Kosower \cite{Bern:1990cu}, and M.\ J.\ Strassler \cite{Strassler:1992zr}, the worldline formalism has developed into a useful method for computing scattering amplitudes and effective actions in quantum field theory \cite{Schubert:2001he}. In particular, this formalism is suitable for studying the effects of quantum fluctuations of matter fields on curved spacetimes. First, a worldline representation for the effective action on an arbitrary gravitational background has been set up for scalar \cite{Bastianelli:2002fv} and fermionic \cite{Bastianelli:2002qw} fields. Later, the case of a vector field has been addressed in \cite{Bastianelli:2005vk}, where -- more generally -- the first Seeley-DeWitt coefficients for an antisymmetric tensor of arbitrary rank have been computed. Seeley-DeWitt coefficients for higher-spin fields on conformally flat manifolds have also been computed with worldline techniques in \cite{Bastianelli:2008nm,Corradini:2010ia,Bastianelli:2012bn}. The worldline formalism is now established as a very efficient tool for quantum field theory computations, in particular, for the study of anomalies (see \cite{Bastianelli:2006rx} and references therein).

In spite of the broad applications developed so far, field theories with boundaries have been more elusive: A worldline formulation which allows analytical computations for quantum fields on manifolds with boundaries is still not known. In fact, the worldline representation for a field $\Phi(x)$ on a spacetime $M$ is based on the correspondence between the spectral modes of quantum fluctuations of $\Phi(x)$ and the Hamiltonian of an auxiliary point-particle with target space $M$. Thus, since the one-loop effective action of the field theory is given by the first quantization of the point-particle, one is lead to consider its path integral over the set of closed trajectories $x^\mu(\tau)$ on $M$\,\footnote{In this formalism, the tree level propagator has a similar representation in terms of a path integral over open lines.}. Consequently, if the manifold has boundary $\partial M$, one expects the path integration domain to be restricted in accordance with the specific conditions on $\Phi(x)$ at $\partial M$. For example, if the quantum field is subject to Dirichlet boundary conditions at $\partial M$ then the path integration must be performed over those closed worldlines $x^\mu(\tau)$ on $M$ which do not intersect $\partial M$~\footnote{Indeed, rigid boundaries can be modeled by coupling $\Phi^2(x)$ to a delta-function with support on $\partial M$ and taking the coupling constant $\lambda\to\infty$. In this limit, the contribution of worldlines which intersect the boundary gets exponentially suppressed.}. However, worldline techniques to perform this restriction on the path integral domain have not been devised yet. The purpose of the present article is to put forth a procedure that can be applied to certain geometries.

The first question is to write down a path integral quantization for a point-particle on a bounded region. Roughly, the main difficulty is that the Gaussian measure (and its moments) can be easily integrated on $\mathbb{R}^D$ but not on its bounded subsets. Dirichlet boundary conditions on a $(D-1)$-di\-men\-sion\-al surface $\Sigma$ can be modeled on the whole $\mathbb{R}^D$ through the coupling $\lambda\,\delta_\Sigma(x)$ to a delta-function with support on $\Sigma$: in the limit of infinite coupling $\lambda\to\infty$ one reproduces Dirichlet conditions. This approach was introduced in the worldline context in \cite{Gies:2003cv} (for a similar mechanism for Neumann boundary conditions, see \cite{Fosco:2009ic}). However, usual analytic worldline techniques require to treat interaction terms perturbatively; such procedure would thus lead to an expansion in positive powers of $\lambda$, and the limit $\lambda\to\infty$ appears ill-defined.

In the context of infinite flat walls, M.~S.~Marinov proposed a different analysis in terms of nontrivial topology in phase space \cite{Marinov:1980vn} but this has not been pursued in the worldline formalism. A different approach which might be adequate for a worldline formulation is given by I.\ S\"okmen who solved the path integral for a particle inside an infinite rectangular well by performing a canonical transformation that takes the particle to the whole line under a Rosen-Morse potential \cite{iS}. However, this solution strongly relies on a particular transformation which holds in this specific one-dimensional setting.

A concrete application of the worldline formalism in the presence of a boundary has been given in \cite{Bastianelli:2006hq,Bastianelli:2008vh}, where image charges have been used to compute the Seeley-DeWitt coefficients for a scalar quantum field on the $D$-dimensional half-space limited by an infinite flat hyperplane. However, the method of images is only applicable to flat boundaries\footnote{Note that although charge images are used for the Laplace equation with spherical boundaries, the same procedure does not work for the heat equation, which describes time evolution for the auxiliary particle.} so to deal with more general cases one needs to introduce a different technique.

In the present work we show how to apply worldline techniques to study a quantum field confined to the $D$-dimensional ball $B^D$ under both Dirichlet and Neumann conditions on the spherical boundary $S^{D-1}$. The procedure, which singles out in the path integral the contributions of worldlines which reach the boundary from those which lie entirely in the bulk, allows one to determine the heat-trace asymptotics of the Laplacian on the compact region $B^D$. Off-diagonal elements of the heat-kernel could likewise be determined were we to
exchange the closed paths for open lines with endpoints at the spatial points in question.

Our first step is to conformally project the compact flat manifold $B^D$ onto the half-space $\mathbb{R}^+\times \mathbb{R}^{D-1}$, which then acquires a non-trivial (but flat) induced metric. In this way, the boundary $S^{D-1}$ is mapped onto the $(D-1)$-dimensional hyperplane. The next step is to duplicate the image of $B^D$ to build up another region $\tilde{B}^D\approx \mathbb{R}^D$ by reflecting the half-space through its boundary and endowing the resulting full space with the symmetric extension of the original induced metric (see e.g.\ Figure 1 in Section \ref{geom}). The region $\tilde{B}^D$ is no longer flat because the symmetric extension introduces a Heaviside-function on the metric, which is thus non-smooth at the interface $\mathbb{R}^D$. Besides, path integration of a point-like particle on curved space corresponds to a $0+1$ sigma model which requires certain counterterms -- specific to each regularization -- that are necessary to maintain general coordinate invariance \cite{Bastianelli:2006rx}. In particular, the counterterm required by time-slicing renormalization contains a term proportional to the curvature of $\tilde{B}^D$, which is given by a delta-function with support at the interface. As a consequence, the computation of the heat-trace in these coordinates amounts to obtaining the point-particle expectation values of combinations of delta- and Heaviside-functions.

It is important to remark that under the conformal map the metric of the half-space is an inverse polynomial so we use the worldline formulation in phase space. Moreover, we use image charges to separate ``direct'' and ``indirect'' contributions to the transition amplitude, according to whether the end-point of the trajectory lies in the physical region $B^D\subset\tilde{B}^D$ or not. Finally, we illustrate the whole procedure in $D=2$ by computing the leading direct and indirect contributions which correspond to the volumes of the disc $B^2$ and its boundary $S^{1}$, as well as the next-to-leading contribution to obtain the Seeley-DeWitt coefficient $a_2$ which gives the trace anomaly.

The organization of the article is as follows. In Section \ref{effactscafie} we give an example of the relation between the heat-trace and effective actions in quantum field theories. In Section \ref{geom} we describe the construction of $\tilde B^D$ as gluing two copies of the ball $B^D$ along its boundary. We compute its geometric properties and define a convenient splitting of its metric into its smooth and singular parts. Next, in Section \ref{tra-amp} we use path integrals in phase space to compute the transition amplitude of a point-particle in the curved background $\tilde B^D$. Section \ref{main} contains the main result of this article, where we use the path integral expression of the previous section to write down a worldline realization of the heat-trace of the Laplacian on the $D$-dimensional ball $B^D$. Both Dirichlet and Neumann boundary conditions are considered. We also give the expressions for the two-point functions (in the worldline) which permit the perturbative evaluation of the path integral for small values of the (Euclidean) proper time; this gives the Seeley-DeWitt coefficients of the corresponding field theory. The procedure to compute the heat-trace asymptotic expansion is depicted in Section \ref{SDW coeff} for the first few Seeley-DeWitt coefficients in the two-dimensional case. Finally, in Section \ref{conclu} we draw some considerations on the applications of our results. In particular, we discuss in some detail the possibility to implement more general boundary conditions. In addition, we comment on the eventual use of our worldline representation in numerical computations in quantum field theory. Some complementary calculations are left to the appendices.

\section{Effective action}\label{effactscafie}

Let us consider a free real scalar field $\varphi(t,x)$ of mass $m$ confined to a spacelike $D$-dimensional manifold $x\in M$, and minimally coupled to gravity. The Euclidean action reads
\begin{align}
	S[\varphi]=\int_{\mathbb{R}\times M}dt\,dx\,\sqrt{g}
	\ \left\{ \frac12\,(\partial\varphi)^2
	+\frac12\,m^2\ \varphi^2\right\}\,,
\end{align}
where $g$ is the determinant of the metric in $M$. The effective action up to one-loop order is
\begin{align}
	\Gamma[\phi]&=S[\phi]-\hbar\log\int\mathcal D\varphi\ e^{-S[\varphi]}\\[2mm]
	&=S[\phi]+\frac{\hbar}2\log{\rm Det}\, \left(-\partial^2_t-\triangle+m^2\right)\,,
\end{align}
where $\triangle$ is the Laplacian on $M$. The identity
\begin{align}
	\log{\lambda}=-\int_0^\infty \frac{dT}{T}\ \left(e^{-T\lambda}-e^{-T}\right)
\end{align}
($\lambda$ is interpreted as an eigenvalue of the operator $-\partial^2_t-\triangle+m^2$) motivates Schwinger's proper-time regularization, which represents the divergent functional determinant in terms of the heat-trace of the Laplacian,
\begin{align}\label{ea}
	\Gamma[\varphi]=S[\varphi]-\frac{\hbar}2
	\int_0^\infty \frac{dT}{T}\ \frac{e^{-Tm^2}}{\sqrt{4\pi T}}\ {\rm Tr}\, e^{-T\left(-\triangle\right)}\,.
\end{align}
Under quite general conditions the heat-trace of the Laplacian $\triangle$ admits the following short time asymptotic expansion \cite{Gilkey:1995mj},
\begin{align}\label{heat-exp}
{\rm Tr}\,e^{-T(-\triangle)}
&\sim \frac{1}{(4\pi T)^{\frac{D}2}}\ \sum_{n=0}^\infty \ a_n(M)\ T^\frac{n}2\,,
\end{align}
where the Seeley-DeWitt coefficients $a_n(M)$ can be computed in terms of geometric invariants of $M$ and its boundary $\partial M$. The use of this expansion in \eqref{ea} shows that the coefficients $a_n(M)$ with $0\leq n\leq D+1$ give the one-loop divergences of the effective action. In particular, the first two coefficients are given only by the volumes of the manifold and its boundary, $a_0(M)\sim {\rm Vol}\,(M)$ and $a_1(M)\sim {\rm Vol}\,(\partial M)$, and do not depend on any other geometric property of space \cite{Vassilevich:2003xt}. On the other hand, the coefficients $a_n(M)$ with $n>D+1$ contribute to the one-loop effective action.

This is just one example of the applications of heat-kernel techniques to the perturbative study of quantum field theories. In this article we will show how to compute the coefficients $a_n(B^D)$ (i.e.\ the heat-trace expansion \eqref{heat-exp} in the compact region inside $S^{D-1}$) using worldline techniques.

\section{Geometry of $\tilde{B}^D$}\label{geom}

As we will show in Section \ref{tra-amp}, the heat-trace asymptotics \eqref{heat-exp} for the Laplacian on the $D$-dimensional ball $B^D$ is determined by the path integral over closed trajectories of a non-relativistic particle. In order to study the dynamics of this particle, it is convenient to identify $B^D$ with a $D$-dimensional half-space which we then embed into a whole space $\mathbb{R}^D$, denoted $\tilde{B}^D$, that represents two copies of the original ball $B^D$ glued together along the interface $\partial B^D\approx S^{D-1}$ as in Figure 1.

\begin{figure}\label{fig}
	\centering
	\includegraphics[width=1\textwidth]{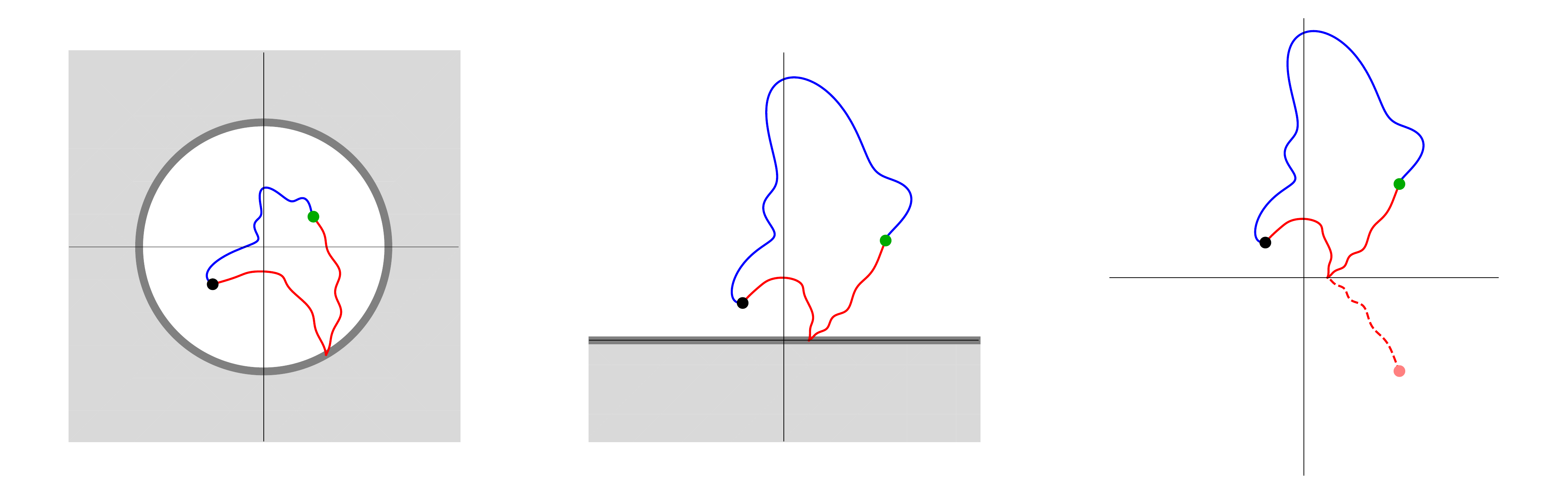}
	\caption{\small Worldlines in $D=2$. The first figure from the left shows two worldlines inside the disc $B^2$ from the initial point {\color{black}$\bullet$} to the final point {\color{ForestGreen}$\bullet$}: the blue trajectory lies entirely within the bulk; the red one hits the boundary once. The second picture displays the same elements after a conformal transformation. The last picture represents $\tilde B^2$: The boundary turns into an interface. A typical curve that hits the boundary (from {\color{black}$\bullet$} to {\color{ForestGreen}$\bullet$}) has a corresponding curve (from {\color{black}$\bullet$} to {\color{pink}$\bullet$}) where the last segment from the boundary to the end-point is switched to its reflection with respect to the interface. The contribution of the original curve is called ``direct''; the second, ``indirect''.}
\end{figure}

Let us begin by considering $B^D$, the interior of $S^{D-1}$, described by coordinates $y=(y_1,\ldots,y_D)\in\mathbb{R}^D$ such that $y^2=y_iy_i\leq 1$. We regard $B^D$ as a flat manifold. Next, we define the following variables $x=(x_1,\ldots,x_D)\in\mathbb{R}^D$,
\begin{align}
	x_i&=\frac{2y_i}{1+y^2-2y_D}\qquad {\rm for\ }i=1,\ldots,D-1\,,\\[2mm]
	x_D&=\frac{1-y^2}{1+y^2-2y_D}\,.
\end{align}
This conformal transformation maps the ball $B^D$ onto the upper half-space $\mathbb{R}^{D-1}\times \mathbb{R}^+$ (to which we also refer as $B^D$) described by $x_D\geq 0$. The original boundary $S^{D-1}$ is mapped onto the horizontal hyperplane $x_D=0$. The north pole $y=(0,\ldots,0,1)$ is mapped to infinity.

The induced (flat) metric in $x$-coordinates is
\begin{align}
  ds^2=\frac{4}{\left(1+x^2+2x_D\right)^2}\ dx_idx_i\,.
\end{align}
Note that the metric is an inverse polynomial. Finally, we extend this metric to the whole $\mathbb{R}^D$ by making a symmetric reflection with respect to the hyperplane $x_D=0$. This ``doubled ball'', which we denote $\tilde B^D$, has metric
\begin{align}\label{dd}
  g_{ij}=\frac{4}{\left(1+x^2+2|x_D|\right)^2}\ \delta_{ij}\,.
\end{align}
Note now that the metric is no longer analytic for it has a discontinuous normal derivative at the fixed points $x_D=0$. The corresponding integration measure is
\begin{align}
	\sqrt {g}=\frac{2^D}{\left(1+x^2+2|x_D|\right)^D}\,.
\label{eq:rg}
\end{align}
The Christoffel symbols are
\begin{align}
  \Gamma^k_{ij} ={\sqrt{g}}^{\frac{1}{D}}\,
  \left\{
  \delta_{ij}x_k - \delta_{ik}x_j - \delta_{jk}x_i
  +\epsilon(x_D)\,(\delta_{ij}\delta_{kD} - \delta_{ki}\delta_{jD} - \delta_{jk}\delta_{iD})
  \right\}\,,
\end{align}
where $\epsilon(x_D)=\pm 1$ for $x_D\gtrless 0$; thus, $\Gamma_{ij}^k$ is in general discontinuous at the boundary. The Ricci tensor and scalar are
\begin{align}
  R_{ij}&=4\ \frac{\delta_{ij}+(D-2)\delta_{iD}\delta_{jD}}{1+x^2}\ \delta(x_D)\,,
  \\[2mm]
  R&=2(D-1)\left(1+x^2\right)\,\delta(x_D)\,,
\label{eq:R}
\end{align}
which have support only at the boundary. The manifold $\tilde B^D$ is therefore not flat.

Path integration in curved spacetimes requires the introduction of an additional counterterm potential to ensure coordinate invariance. In the present manuscript we find it convenient, as it will be clear below, to expand the inverse metric, which is coupled to particle momenta. Thus instead of a configuration space path integral, we will use a phase-space path integral which, in curved space, is suitably described in terms of the Time Slicing formulation developed in~\cite{DeBoer:1995hv} (for a review see~\cite{Bastianelli:2006rx}), which involves the potential:
\begin{align}\label{counterterm}
  &\Delta H_{\rm TS}(x)=-\frac{1}{4}\,R+\frac{1}{4}\,g^{ij}\,\Gamma^\ell_{ik}\Gamma^k_{j\ell}\,,\nonumber\\[2mm]
  &=-\frac{D-1}{2}\, \left(1+x^2\right) \delta(x_D) 
  -\frac{D-2}{4}\, \left(1+x^2+2|x_D|\right)\,.
\end{align}
It is now convenient to separate the analytic from the non-analytic part of the metric. To do that we write the inverse metric as
\begin{align}\label{sepginv}
  g^{ij}=\left\{h(x)-f(x)\,\theta(-x_D)\right\}\,\delta_{ij}\,,
\end{align}
where $\theta(-x_D)$ is the Heaviside-function and
\begin{align}
  h(x)&=\frac{1}{4}\,\left(1+x^2+2x_D\right)^2\,,\\
  f(x)&=2x_D\left(1+x^2\right)\,,
\end{align}
are smooth functions in the whole space $\mathbb{R}^D$. Note that one could have instead separated the ``smooth'' part of the inverse metric differently, e.g.\ using the sign-function $\epsilon(x_D)$; the convenience of leaving the smooth part $h(x)$ equal to the original inverse metric on $B^D$ lies in the fact that with this choice purely bulk contributions vanish. Also, had we used the complementary Heaviside-function $\theta(x_D)$, we would have obtained a smooth part with a singularity at the point $x=(0,\ldots,0,1)\in B^D$ (the center of the ball). Since the heat-trace involves an integration over the ``physical'' region $x_D\geq0$ only, a singularity at the image point $\tilde x=(0,\ldots,0,-1)$ is innocuous.

\section{Transition amplitudes in $\tilde B^D$}\label{tra-amp}

In this section we study the transition amplitudes of a non-relativistic particle on the manifold $\tilde B^D$ by considering all trajectories $x(\tau)$ which go from the point $x\in\mathbb{R}^D$ to the point $x'=x+\xi\in\mathbb{R}^D$ in Euclidean time $T$. For convenience, we describe these trajectories as $x(\tau)=x_0(\tau)+q(\tau)$, where $x_0(\tau)$ is the straight line
\begin{align}
  x_0(\tau)=\xi\,\frac{\tau}{T}+x
\end{align}
that connects the point $x_0(0)=x$ with $x_0(T)=x'$, and $q(\tau)$ representing quantum fluctuations under homogeneous Dirichlet conditions $q(0)=q(T)=0$.

The transition amplitude can then be represented in terms of the phase-space path integral \cite{Bastianelli:2006rx}
\begin{align}
  \langle x'| e^{-T(-\triangle)} |x\rangle
  &=\left(\sqrt{g(x')}\sqrt{g(x)}\right)^{-\frac12}\times\mbox{}\nonumber\\[2mm]
  &\mbox{}\times\int\mathcal Dp\mathcal Dq
  \ e^{-\int_0^Td\tau\,\left\{g^{ij}\,p_ip_j-i p_i\left(\frac{\xi_i}T+\dot q_i\right)
  +\Delta H_{\rm TS}\right\}}\,.
\end{align}
In the integrand both $g^{ij}$ and $\Delta H_{\rm TS}$ are evaluated at $x_0(\tau)+q(\tau)$.

In order to keep track of the different powers of the (small) variable $T$, we turn to dimensionless quantities: $\tau\to T\tau$, $p(\tau)\to p(\tau)/\sqrt{T}$ and $q(\tau)\to \sqrt{T}\,q(\tau)$. Next, we use \eqref{sepginv} and make an expansion around the fixed initial point $x$ to separate quadratic terms from interaction terms (and shift the momentum variables as $p\to p+i\frac{\xi}{2h(x)\sqrt T}$). We thus obtain
\begin{align}\label{ta}
\langle x'| e^{-T(-\triangle)} |x\rangle
&=\left(\sqrt{g(x')}\sqrt{g(x)}\right)^{-\frac12}
\ e^{-\frac{\xi^2}{4Th(x)}}\times\mbox{}\nonumber\\[2mm]
&\mbox{}\times\int\mathcal Dp\mathcal Dq
\ e^{-\int_0^1d\tau\,\left\{h(x)\,p^2-i p\dot q\right\}}
\ e^{-\int_0^1d\tau\ H_{\rm int}(\tau)}\nonumber\\[2mm]
&=\left(\sqrt{g(x')}\sqrt{g(x)}\right)^{-\frac12}
\ e^{-\frac{\xi^2}{4Th(x)}}
\ \left\langle e^{-\int_0^1d\tau\ H_{\rm int}(\tau)}\right\rangle\,,
\end{align}
where the interaction Hamiltonian is
\begin{align}\label{ta2}
  &H_{\rm int}(\tau)=
  -T\ \frac{D-1}{2}\left[1+(x+\tau\xi+\sqrt T q)^2\right]\,\delta(x_D+\tau\xi_D+\sqrt Tq_D)\nonumber\\[2mm]
  &\mbox{}-T\ \frac{D-2}{4}\left[1+(x+\tau\xi+\sqrt T q)^2+2\left|x_D+\tau\xi_D+\sqrt Tq_D\right|\right]\nonumber\\[2mm]
  &\mbox{}+\left(p+i\tfrac {\xi}{2h(x)\sqrt{T}}\right)^2
  \ \bigg\{-f(x+\tau\xi+\sqrt T q)\,\theta(-x_D-\tau\xi_D-\sqrt T q_D)
  \nonumber\\[2mm]
  &\mbox{}+\partial_ih(x)\,(\tau\xi_i+\sqrt T q_i)
  +\frac12\partial^2_{ij}h(x)\,(\tau\xi_i+\sqrt T q_i)(\tau\xi_j+\sqrt T q_j)
  +\ldots\bigg\}\,,
\end{align}
where the last line simply is $h(x(\tau)) -h(x)$, expressed as a Taylor expansion about the point $x$.

The expectation value in \eqref{ta} represents the phase-space integration over trajectories in configuration space $q(\tau)$ which satisfy homogeneous Dirichlet conditions $q(0)=q(1)=0$, and completely free trajectories $p(\tau)$ on which no boundary conditions are imposed. This makes the relevant quadratic operator in the Gaussian measure of \eqref{ta} invertible. It is important to remark that due to the rescalings in the phase-space worldlines the expectation value $\langle 1\rangle$ does depend on $T$, although it is at this point not manifest in the notation. In the next section, we will restore the dependence on $T$ to compute the correct normalization.

Each of the four lines in \eqref{ta2} corresponds to one of four types of interaction terms: the first two arising from the time-slicing counterterm (which we will call $\delta$-term and $\epsilon$-term, correspondingly), the third one corresponding to the discontinuity introduced by the reflection of the metric along the boundary (we will call it $\theta$-term) and the fourth, to the Taylor expansion of the metric around the fixed point $x$ (these will be called $h$-terms).

\section{The heat-trace in $B^D$}\label{main}

This section summarizes the main result of this article: a worldline representation for the heat-trace of the Laplacian on $B^D$. We show how to use the transition amplitude \eqref{ta} to determine the asymptotic expansion through a perturbative calculation in the worldline.

According to the image charge method, the transition amplitude from an initial point $x=(x_1,\ldots,x_D)$ to another point $x'=(x'_1,\ldots,x'_D)$ in $B^D$ under Dirichlet or Neumann boundary conditions can be obtained by subtracting or adding the transition amplitude in $\tilde B^D$ from the same initial point to an image end-point at $\tilde x=(x'_1,\ldots,x'_{D-1},-x'_D)$. As we have already mentioned, the transition amplitude in the half-space $B^D$ under Dirichlet boundary conditions is given by the contributions of all paths from $x$ to $x'$ which do not intersect the boundary; these contributions can be obtained from all paths in the whole space $\tilde B^D$ from $x$ to $x'$ if we subtract the extra contributions from paths that hit the boundary, but these are equivalent to the contributions of all paths from $x$ to $\tilde x$  (see Figure 1 in Section \ref{geom}). A similar argument can be drawn for Neumann boundary conditions.

For the heat-trace, this leads to
\begin{align}\label{images}
{\rm Tr}\,e^{-T(-\triangle)}&=
\int_{\mathbb{R}^{D-1}\times\mathbb{R}^+}dx\sqrt{g}
\ \langle x| e^{-T(-\triangle)} |x\rangle
\mp\int_{\mathbb{R}^{D-1}\times\mathbb{R}^+}dx\sqrt{g}
\ \langle \tilde x| e^{-T(-\triangle)} |x\rangle\,,
\end{align}
for Dirichlet and Neumann boundary conditions, respectively. We refer to the first term in the r.h.s.\ as the direct contribution, whereas the second term is referred to as the indirect contribution. The transition amplitudes in this expression correspond to trajectories in $\tilde B^D$ so they can be computed using \eqref{ta}. The operator $H_{\rm int}(\tau)$ given by \eqref{ta2} must be evaluated at $\xi=0$ for direct contributions, and at $\xi=(0,\ldots,0,-2x_D)$ for indirect ones. We now describe how to perform a perturbative expansion of the expectation value in \eqref{ta} to obtain an asymptotic expression of the heat-trace. 

In general, to compute $p$- and $q$-correlators with respect to the expectation value in \eqref{ta} one introduces two-component external sources $k(\tau),j(\tau)$ and define the (free) generating functional
\begin{align}\label{gen}
&Z[k,j]=\left\langle e^{i\int_0^1d\tau\,(kp+jq)} \right\rangle
=\int\mathcal Dp\mathcal Dq
\ e^{-\int_0^1d\tau\,\left\{h(x)\,p^2-i p\dot q\right\}
+i\int_0^1d\tau\,\left(kp+jq\right)}
\nonumber\\[2mm]
&=e^{-\frac12\int d\tau d\tau'\,\left\{\frac{1}{2h(x)}\,k(\tau)k(\tau')
	+h(x)\,G(\tau,\tau')j(\tau)j(\tau')+i\,\mbox{}^{\bullet}G(\tau,\tau')k(\tau)j(\tau')\right\}}
\ \langle 1\rangle\,,
\end{align}
where
\begin{align}
G(\tau,\tau')&=-|\tau-\tau'|-2\tau\tau'+\tau+\tau'\,,
\label{eq:DBC-Green}\\
\mbox{}^\bullet G(\tau,\tau')&=-\epsilon(\tau-\tau')-2\tau'+1
\end{align}
are Green functions given by the matrix elements of the inverse of the quadratic operator in the Gaussian measure. The notation $\mbox{}^\bullet G(\tau,\tau')$ points out that this function is the derivative of $G(\tau,\tau')$ with respect to its first argument.

Two-point functions can be computed as the functional derivatives of \eqref{gen},
\begin{align}
\langle p_i(\tau)p_j(\tau')\rangle&
=\langle 1\rangle\, \delta_{ij}\,\frac{1}{2h(x)}\,,\label{prop1}\\[2mm]
\langle q_i(\tau)q_j(\tau')\rangle&
=\langle 1\rangle\, \delta_{ij}\,h(x)\,G(\tau,\tau')\,,\label{prop2}\\[2mm]
\langle p_i(\tau)q_j(\tau')\rangle&
=\langle 1\rangle\, \delta_{ij}\,\frac{i}{2}\ \mbox{}^{\bullet}G(\tau,\tau')\,.\label{prop3}
\end{align}
Other useful correlation functions are summarized in Appendix \ref{corre}. Due to the boundary conditions on the phase-space trajectories $q(\tau)$ and $p(\tau)$, the normalization\footnote{Note that we have undone the rescalings to restore the dependence on $T$.}
\begin{align}
	\langle 1\rangle=\int\mathcal Dp\mathcal Dq
	\ e^{-\int_0^Td\tau\,\left\{h(x)\,p^2-i p\dot q\right\}}
\end{align}
can be regarded as the transition amplitude in flat space of a free particle of mass $(2h(x))^{-1}$ with initial and final points at the origin. This is simply given by
\begin{align}
\langle1\rangle=\frac1{(4\pi T h(x))^{\frac{D}{2}}}\,.
\end{align}
The expectation value in \eqref{ta} can now be computed by expanding the exponential and evaluating the expectation values of the different terms in $H_{\rm int}$ given by \eqref{ta2}. These terms contain powers of the canonical operators $p,q$ whose $n$-point functions can be obtained using Wick's theorem and expressions \eqref{prop1}-\eqref{prop3}. However, expression \eqref{ta2} also contains delta- and theta-functions. In the next section we show how to handle these distributions with worldline techniques and illustrate the whole procedure by computing the first three Seeley-DeWitt coefficients. For simplicity we consider the case $D=2$ -- note that in this case the $\Gamma\Gamma$ term from \eqref{counterterm} identically vanishes.

\section{The trace anomaly in $B^2$}\label{SDW coeff}

To clarify the procedure described in the previous section, we will now use equation \eqref{images} -- together with \eqref{ta} -- to reproduce in full detail the first Seeley-DeWitt coefficients $a_0(B^2),a_1(B^2),a_2(B^2)$ on the disc. We consider direct and indirect contributions separately.

\subsection{Direct contributions}

To obtain the direct contribution to the heat-trace we evaluate expressions \eqref{ta} and \eqref{ta2} for $\xi=0$ and integrate over $B^2$,
\begin{align}\label{dir}
  \int_{\mathbb{R} \times \mathbb{R}^+} dx \sqrt{g}
  \ \langle x| e^{-T(-\triangle)} |x\rangle
  =\int_{\mathbb{R} \times \mathbb{R}^+} dx\ \left\langle e^{-\int_0^1 d\tau\ H^{\rm dir}_{\rm int}(\tau)}\right\rangle\,,
\end{align}
where
\begin{align}\label{LintD}
H^{\rm dir}_{\rm int}(\tau)
&=-\frac T2\left[1+(x_1+\sqrt T q_1)^2\right]
\,\delta(x_2+\sqrt Tq_2)+\mbox{}\nonumber\\
&\mbox{}-p^2\,f(x+\sqrt T q)\,\theta(-x_2-\sqrt T q_2)
+\mbox{}\nonumber\\
&\mbox{}+p^2\ \left\{\partial_ih(x)\, q_i\,\sqrt T
+\frac12\partial^2_{ij}h(x)\,q_iq_j\,T
+\ldots\right\}\,.
\end{align}
Note that since $x=x'$ then $g(x)=g(x')$ and there is no measure $\sqrt g$ in the r.h.s.\ of \eqref{dir}.

The leading contribution to \eqref{dir} can be straightforwardly computed as
\begin{align}
\int dx\ \left\langle 1\right\rangle
=\int dx\ \frac1{4\pi Th(x)}=\frac{1}{4T}\,,
\end{align}
which according to \eqref{heat-exp} gives the well-known result for the first Seeley-DeWitt coefficient $a_0(B^2)={\rm Vol}(B^2)=\pi$ \cite{Vassilevich:2003xt}.

Let us now compute the next-to-leading direct contribution which comes from all three lines in the r.h.s.\ of \eqref{LintD}, namely, the $\delta$-term, $\theta$-term and $h$-terms. We begin with the leading contribution from the third line in \eqref{LintD}, i.e., the $h$-terms,
\begin{align}
  &\int dx\ \bigg\{
  -\frac T2\,\partial^2_{ij}h(x)\int_0^1d\tau\,\langle p^2(\tau)q_i(\tau)q_j(\tau)\rangle+\mbox{}\nonumber\\[2mm]
  &\mbox{}+\frac T2\,\partial_{i}h(x)\partial_{j}h(x)\int_0^1d\tau\int_0^1d\tau'
  \,\langle p^2(\tau)p^2(\tau')q_i(\tau)q_j(\tau')\rangle\bigg\}\,.
\end{align}
The four- and six-point functions in this expression can be computed using the free propagators \eqref{prop1}-\eqref{prop3},
\begin{align}
\langle p^2(\tau)q_i(\tau)q_j(\tau)\rangle&
=\langle 1\rangle\, \delta_{ij}\,\frac12\left(2G(\tau,\tau)-\mbox{}^{\bullet}G(\tau,\tau)^2\right)\,,\\[2mm]
\langle p^2(\tau)p^2(\tau')q_i(\tau)q_j(\tau')\rangle&
=\langle 1\rangle\, \delta_{ij}\,\frac1{2h(x)}\,
\big\{ 4G(\tau,\tau')+\mbox{}\nonumber\\[1mm]
&\mbox{}-\mbox{}^{\bullet}G(\tau',\tau')\mbox{}^{\bullet}G(\tau',\tau)-\mbox{}^{\bullet}G(\tau,\tau)\mbox{}^{\bullet}G(\tau,\tau')+\mbox{}\nonumber\\[1mm]
&\mbox{}-\mbox{}^{\bullet}G(\tau,\tau)\mbox{}^{\bullet}G(\tau',\tau')-\mbox{}^{\bullet}G(\tau,\tau')\mbox{}^{\bullet}G(\tau',\tau)\big\}
\,.
\end{align}
Thus, the leading direct contribution of the $h$-terms is
\begin{align}
\frac{T}{12}\int dx\ \langle 1\rangle\left\{
-\partial^2h(x)\ +\partial_{i}h(x)\partial_{i}h(x)\,\frac{1}{h(x)}\right\}=0\,.
\end{align}
This cancellation is to be expected because $h$-terms are not related to the singularities at the interface $x_2=0$ but only to the smooth part of the inverse metric; since the disc is flat no purely bulk contributions should appear.

Let us next study the leading contribution of the first line in the r.h.s.\ of \eqref{LintD}, i.e., the $\delta$-term,
\begin{align}\label{dirdelta}
  &\frac T2\int dx\ \int_0^1d\tau
  \left\langle\left[1+\left(x_1+\sqrt T q_1(\tau)\right)^2\right]\,\delta\left(x_2+\sqrt T q_2(\tau)\right)\right\rangle\sim\nonumber\\
  &\sim\frac T2\int_{\mathbb{R}\times\mathbb{R}^+} dx_1dx_2
  \ \left(1+x_1^2\right)\int_0^1d\tau
  \int_{-\infty}^{\infty}\frac{d\omega}{2\pi}\,e^{i\omega x_2}
  \left\langle e^{i\omega\sqrt T q_2(\tau)}\right\rangle\,,
\end{align}
where we have only retained the leading contribution in $T$, and we have also used the integral representation\footnote{Correlators involving $\delta$-functions of this form have previously been treated in this way in the context of contact interactions between strings \cite{Edwards:2014cga,Edwards:2014xfa} and particles \cite{Edwards:2015hka}.}
\begin{align}
\delta(x)=\int_{-\infty}^\infty \frac{d\omega}{2\pi}\ e^{i\omega x}\,.
\end{align}
The expectation value in \eqref{dirdelta} is given by the generating functional \eqref{gen} for $k_1(\tau')=k_2(\tau')=j_1(\tau')=0$ and $j_2(\tau')=\omega\sqrt T\delta(\tau'-\tau)$,
\begin{align}
\left\langle e^{i\omega\sqrt T q_2(\tau)}\right\rangle
	=e^{-\frac{T}2\omega^2h(x)\,G}
	\ \langle 1\rangle\,;
\end{align}
throughout this article, whenever it becomes clear from the context, we will use the notation $G=G(\tau,\tau)$, as well as $\mbox{}^{\bullet}G=\mbox{}^{\bullet}G(\tau,\tau)$. The leading contribution of the delta-function then results
\begin{align}
&\frac{1}{16\pi^2}\int_{\mathbb{R}\times\mathbb{R}^+} dx_1dx_2
\ \frac{1+x_1^2}{h(x)}\int_0^1d\tau
\int_{-\infty}^{\infty}d\omega\,e^{i\omega x_2}
e^{-\frac{T}2h(x)G\,\omega^2}\sim\frac14\,.
\end{align}
An important remark is now in order: to compute the integration in $x_2$ we have first rescaled $x_2\to\sqrt T x_2$ and, consequently, neglected the dependence of $h(x)$ on $x_2$ to leading order in $T$. This allows a further rescaling of $x_2$ which makes the integrations in $x_2$ and $\tau$ straightforward. These consecutive rescalings of the $x_2$-coordinate will be frequent in the subsequent calculations and can further be justified by first integrating over $\omega$; in the limit $T \rightarrow 0$, the result yields a representation of $\delta(x_{2})$ plus higher-order corrections. To compute subleading contributions one should Taylor expand $h(x)$ around $x_2=0$ and perform consistent calculations order by order in $T$.

Worldline representations admit a variety of approaches. For instance, we could have dealt with the leading contribution of the expectation value \eqref{dirdelta} by neglecting $O(\sqrt T)$ terms in the formal expansion for small $\sqrt T q(\tau)$,
\begin{align}
	\frac T2\int dx\int_0^1d\tau
	\,\left(1+x_1^2\right)\,\delta(x_2)
	\left\langle 1\right\rangle=\frac14\,.
\end{align}
Notwithstanding the simplicity of this procedure, we use the more sound Fourier representation of distributions because it is better suited for higher order calculations, where an expansion of $\delta$-functions becomes inappropriate. See another alternative calculation in Appendix \ref{B}.

To conclude our study of the direct terms, we compute the leading contribution of the second line in r.h.s.\ of \eqref{LintD}, i.e., the $\theta$-term,
\begin{align}\label{dirtheta}
&\int dx\int_0^1d\tau
\ \left\langle p^2(\tau)\,f(x+\sqrt T q(\tau))\,\theta(-x_2-\sqrt T q_2(\tau))\right\rangle=\nonumber\\[2mm]
&=\int dx\int_0^1d\tau\int_{-\infty}^\infty \frac{d\omega}{2\pi i}
\ \frac{e^{-i\omega x_2}}{\omega-i\,0}
\ \left\langle p^2\,f(x+\sqrt T q)\ e^{-i\omega\sqrt T q_2}\right\rangle\,.
\end{align}
Note that this time we have used the Fourier transformation
\begin{align}
\theta(x)=\int_{-\infty}^\infty \frac{d\omega}{2\pi i}\ \frac{e^{i\omega x}}{\omega-i\,0}\,,
\end{align}
which will eventually be complemented with
\begin{align}\label{pv}
	\frac{1}{\omega-i\,0}={\rm P}\left(\frac1\omega\right)+i\pi\delta(\omega)\,.
\end{align}
Upon the rescalings $\omega\to\omega/\sqrt T$ and $x_2\to \sqrt T x_2$, expression \eqref{dirtheta} can be written, to leading order in $T$, as
\begin{align}\label{dirtheta2}
&2T\int dx\ (1+x_1^2)\int_0^1 d\tau\int \frac{d\omega}{2\pi i}
\ \frac{e^{-i\omega x_2}}{\omega-i\,0}
\ \left\langle p^2\,(x_2+q_2)\ e^{-i\omega q_2}\right\rangle\,.
\end{align}
The expectation value will introduce further dependence on $x$ through the function $h(x)$: as before, because of the rescaling in $x_2$, the function $h(x)$ becomes a function of $(x_1,\sqrt T x_2)$ and, to leading order, a function that only depends on $x_1$. To keep this in mind we define $\bar x=(x_1,0)$ and introduce $\bar h=h(\bar x)$.

Let us then compute both expectation values in \eqref{dirtheta2}. Firstly,
\begin{align}\label{pp-exp}
\left\langle p^2(\tau)\,e^{-i \omega q_2(\tau)}\right\rangle&=
-\left.\frac{\delta^2Z[k(\tau'),j(\tau')]}{\delta k_i(\tau)\delta k_i(\tau)}
\right|_{j_2(\tau')=-\omega\,\delta(\tau'-\tau)}\nonumber\\[2mm]
&= \frac{1}{4\pi T h(x)}\left(\frac{1}{h(x)}+\frac14\,\omega^2\,\mbox{}^{\bullet}G^2\right)
e^{-\frac{h(x)}2\omega^2\,G}\,.
\end{align}
Though not explicitly indicated, we have also evaluated the functional derivative at $k_i(\tau')=j_1(\tau')=0$. In a similar fashion we can compute the second expectation value; the result reads
\begin{align}\label{pp-q-exp}
\left\langle p^2\,q_{2}\,e^{-i\omega q_2}\right\rangle
=-\frac{i\omega}{4\pi T h(x)}
\left(G-\frac12\,\mbox{}^{\bullet}G^2+\frac14\,h(x)\omega^2\,G\ \mbox{}^{\bullet}G^2\right)
e^{-\frac{h(x)\,G}2\,\omega^2}\,.
\end{align}
Appendix \ref{corre} contains these and several other correlation functions. Replacing \eqref{pp-exp} and \eqref{pp-q-exp} into \eqref{dirtheta2} we get, to leading order,
\begin{align}
&\int dx\ \frac{1+x_1^2}{2\pi \bar h}\int_0^1 d\tau\int \frac{d\omega}{2\pi i}
\ \frac{e^{-\frac{\bar hG}2\,\omega^2-i\omega x_2}}{\omega-i\,0}\times\mbox{}\nonumber\\[2mm]
&\mbox{}\times\left(\frac{x_2}{\bar h}
-i\omega(G-\tfrac12\,\mbox{}^{\bullet}G^2)
+\tfrac14\,\omega^2\,x_2\mbox{}^{\bullet}G^2
-\tfrac14\,i\omega^3 \bar h\,G\ \mbox{}^{\bullet}G^2\right)=-\frac1{12}\,.
\end{align}
To integrate the first term, proportional to $(\omega-i\,0)^{-1}$, we have used relation \eqref{pv}: the integration of the delta function $\delta(\omega)$ is trivial, whereas the principal value function $P(\omega^{-1})$ allows one to retain only the $\sin$-function contained in the imaginary exponential, which makes the integral convergent at $\omega=0$. Integration of the remaining terms is straightforward.

In conclusion, the $O(T^0)$ direct contribution to the heat-trace, stemming from the three lines in \eqref{LintD}, gives
\begin{align}
	0+\frac14-\frac1{12}=\frac16\,.
\end{align}
The present results can as well be obtained employing String Inspired (SI) worldline Green's function, which correspond to identifying the integration point $x$ with the center of mass of the path, so that the quantum fluctuations are periodic but have no center of mass, i.e.\ $\int d\tau\, q(\tau) =0$. This leads to a Poincar\'e invariant Green's function which satisfies $G(\tau,\tau) =1/6$ and ${}^{\bullet}G(\tau,\tau)=0$ -- see e.g.~\cite{Bastianelli:2008vh}. Thus the computation of the direct contributions turns out to be much simpler. However, unlike the Dirichlet Boundary Conditions (DBC) Green's function of~\eqref{eq:DBC-Green}, it is not obvious how to employ SI periodic Green's function for the indirect contributions, where $\xi\neq 0$. In fact, for the integrated indirect contributions, the transversal coordinate $x_D$, unlike the parallel coordinates, becomes anti-periodic in the String Inspired approach. So, one should use the anti-periodic bosonic Green's function for the transversal part, and periodic ones for the parallel coordinates. Here, we prefer to display the results only in terms of the DBC Green's function which applies to both direct and indirect terms in the same way. However, it would be interesting to investigate further on the application of String Inspired Feynman rules to such computations.

\subsection{Indirect contributions}

The indirect contribution to the heat-trace is given by
\begin{align}\label{indir}
  \int_{\mathbb{R} \times \mathbb{R}^+} dx \sqrt{g}
  \ \langle \tilde x| e^{-T(-\triangle)} |x\rangle
  =\int_{\mathbb{R} \times \mathbb{R}^+} dx\ e^{-\frac{x_2^2}{Th(x)}}
  \ \left\langle e^{-\int_0^1d\tau H^{\rm ind}_{\rm int}(\tau)}\right\rangle\,,
\end{align}
with
\begin{align}\label{LintI}
  &H^{\rm ind}_{\rm int}(t)
  =-\frac T2\left[1+\left(x_1+\sqrt Tq_1\right)^2\right]
  \,\delta\left((1-2\tau)\,x_2+\sqrt Tq_2\right)+\mbox{}\nonumber\\
  &\mbox{}+\left(p+i\frac {\xi}{2h\sqrt T}\right)^2
  \ \bigg\{-f\left(x+\tau\,\xi+\sqrt Tq\right)
  \,\theta\left(-(1-2\tau)\,x_2-\sqrt T q_2\right)
  +\mbox{}\nonumber\\
  &\mbox{}+\partial_ih\,\left(\tau\,\xi_i+\sqrt T q_i\right)
  +\frac12\partial^2_{ij}h\,\left(\tau\,\xi_i+\sqrt Tq_i\right)\left(\tau\,\xi_j+\sqrt Tq_j\right)
  +\ldots\bigg\}\,,
\end{align}
where $\xi=(0,-2x_2)$. Note that due to the symmetric extension of the metric $g(x)=g(\tilde x)$ and there is no measure $\sqrt g$ in the r.h.s.\ of \eqref{indir}. 

Let us begin with the leading contribution
\begin{align}
  \int dx\ e^{-\frac{x_2^2}{Th(x)}}\,\langle1\rangle
  =\frac{1}{\pi T}\int dx_1dx_2
  \ \frac{e^{-\frac{4x_2^2}{T[x_1^2+(x_2+1)^2]^2}}}{[x_1^2+(x_2+1)^2]^2}
  =\frac{\sqrt{\pi}}{4\sqrt T}+O(\sqrt T)\,.
\end{align}
As before, we have conveniently rescaled $x_2\to \sqrt Tx_2$. This result reproduces the second Seeley-DeWitt coefficient $a_1(B^2)=\sqrt{4\pi}\,{\rm Vol}(S^1)/4=\pi^\frac32$ \cite{Vassilevich:2003xt}. Note also that this contribution does not contain any $O(T^0)$ term.

Next, we consider the leading contribution of the third line (the $h$-terms) in \eqref{LintI}. However, these are purely bulk terms so, as expected, a straightforward computation shows that this contribution vanishes.

The leading contribution of the $\delta$-term in \eqref{LintI} can be carried out along the same lines as for the direct terms. In fact, one obtains the same result, namely $\frac14$. this can be understood from the fact that a $\delta$-type expectation value $\langle\delta(x(\tau))\rangle$ with support at the boundary must vanish for Dirichlet conditions. Therefore, direct and indirect contributions must coincide.

Finally, we must compute the leading contribution of the $\theta$-term in \eqref{LintI},
\begin{align}
\int dx d\tau\ e^{-\frac{x_2^2}{Th}}\left\langle
\left(p+i\tfrac {\xi}{2h\sqrt T}\right)^2
\,f\left(x+\tau\,\xi+\sqrt Tq\right)
\,\theta\left(-(1-2\tau)\,x_2-\sqrt T q_2\right)
\right\rangle\,.
\end{align}
Computing the correlation functions (see Appendix \ref{corre}) we get
\begin{align}\label{intw}
&-\frac{i}{(2\pi)^2}\int dx d\tau\ e^{-\frac{x_2^2}{\bar h}}
\ \frac{1+x_1^2}{\bar h}
\int d\omega\ \frac{e^{-i\omega(1-2\tau)x_2}}{\omega-i\,0}
\ e^{-\frac{1}{2}\bar h G\,\omega^2}
\times\mbox{}\nonumber\\[2mm]
&\mbox{}\times\left\{
\mbox{}^{\bullet}G\left(2-\frac{x_2^2}{\bar h}\right)\frac{x_2}{\bar h}
-\omega^2\,\mbox{}^{\bullet}G\left(G-\frac14\,\mbox{}^{\bullet}G^2\right)\,x_2
+\mbox{}\right.\nonumber\\[2mm]
&\left.\mbox{}
-i\omega
\left(G-\frac12\,\mbox{}^{\bullet}G^2-(G-\mbox{}^{\bullet}G^2)\,\frac {x_2^2}{\bar h}\right)
-\frac14\,i\omega^3\,G\ \mbox{}^{\bullet}G^2\,\bar h
\right\}=-\frac14\,.
\end{align}
The integrals can be straightforwardly performed by replacing $(\omega-i\,0)^{-1}$ by the principal value ${\rm P}(\omega^{-1})$ (note that the contribution of the delta-function to the first term vanishes due to the integration of $\mbox{}^{\bullet}G$ in the interval $\tau\in[0,1]$).

In conclusion, the $O(T^0)$ indirect contribution, stemming from the three lines in \eqref{LintI}, gives
\begin{align}
0+\frac14-\frac14=0\,.
\end{align}

\subsection{Seeley-DeWitt coefficients}

The results of our calculations give the following leading terms for the heat-trace asymptotics:
\begin{align}\label{real}
  {\rm Tr}\,e^{-T(-\triangle)}
  &\sim \frac{1}{4\pi T}\ \sum_{n=0}^\infty a_n(B^2)\,T^\frac{n}2\nonumber\\[2mm]
  &\sim
  \frac{1}{4T}\mp\frac{\sqrt{\pi}}{4\sqrt T}
  +\frac16
  +O(\sqrt T)\,.
\end{align}
From this expression one reads the Seeley-DeWitt coefficient $a_2=\frac23\pi$ (in agreement with \cite{Vassilevich:2003xt,Kirsten:2001wz}) which gives the integrated two-dimensional trace anomaly
\begin{align}
\int d^2x \sqrt{g}\, \big \langle T^\mu{}_\mu \big\rangle =\frac{1}{4\pi}\, a_2 = \frac16\,.
\end{align}

\section{Conclusions and future work}\label{conclu}

We have considered a scalar field confined to the interior of the sphere $S^{D-1}$ in order to analyze the applicability of a worldline approach to the determination of the heat kernel. By means of an appropriate conformal transformation that maps the spherical boundary into a hyperplane $\mathbb{R}^{D-1}$, the standard usage of image charges allowed us to implement either Dirichlet or Neumann boundary conditions. Together with a symmetric reflection of the metric on the hyperplane, the new coordinates turn the boundary into an interface were a singular curvature develops. Computations of correlation functions in the worldline approach are then implemented in phase space. As an example, we have reproduced for $D=2$ the first three Seeley-DeWitt coefficients.

The procedure that we have set up admits generalizations and concrete applications. Up to now, a worldline formulation of quantum fields on a manifold with boundaries has been established only for a flat boundary: firstly, Dirichlet and Neumann boundary conditions have been analyzed \cite{Bastianelli:2006hq,Bastianelli:2008vh}; later, based on \cite{CMS}, Robin boundary conditions \cite{Bastianelli:2007jr} and specific matching conditions on a flat interface \cite{Vinas:2010ix} were considered. The procedure described in the present article can now be carried out for these more general conditions on a spherical shell and we expect it to grant the familiar calculational efficiency that has previously been seen using first quantized techniques.

The extension to other types of boundary conditions is also worth considering. In \cite{Carreau:1990wh} a Brownian measure which appropriately describes all self-adjoint extensions for a one-dimensional particle between infinite walls has been constructed. This result has been generalized to the infinite dimensional family of boundary conditions for a particle in the half-plane in \cite{Carreau:1991yx}; it would be interesting to consider whether this formulation admits a representation in the worldline formalism (if one aims at considering the most general boundary conditions \cite{Asorey:2004kk}, one must take into account the interesting difficulties discussed in \cite{Asorey:2006ij,Asorey:2007zza}).

In this context it would also be interesting to revisit the detailed analysis in \cite{Graham:2002xq} (see also \cite{Graham:2002fw,Graham:2003ib}) which shows that the Casimir energy in the presence of physical surfaces contains divergencies which cannot be removed by appropriate counterterms in the Lagrangian. These divergencies are innocuous for computing Casimir forces between different surfaces but have cast doubts on the appropriate definition of the Casimir tension on each surface. Actually, the problem appears in the vicinity of delta-like shells, so the worldline formalism could be applied to these models since our approach already incorporates delta-like backgrounds, and can be easily adapted to the computation of the energy densities.

The motivation for these generalizations not only arises from the applications to analytic calculations: in the last two decades, the worldline formalism has proven to provide a powerful tool for the numerical computation of physical quantities in quantum field theory. This approach originated in the numerical evaluation of effective actions in analytically solvable problems \cite{Gies:2001zp}, but since then has been used in a broad variety of contexts. In \cite{Gies:2005bz} it has been used in the Minkowski setting to study pair production by an inhomogeneous background. In particular, worldline numerics is currently one of the few available methods to benchmark experimental measurements on the Casimir effect (first considered in \cite{Gies:2003cv}, see e.g.\ its applications to other geometries in \cite{Schaden:2009zz,Schaden:2011aa}, and to more realistic media in \cite{Mackrory:2016cjx}). In this context, worldline based numerical analysis has been used to determine the range of validity of the scheme known as proximity force approximation \cite{Gies:2006bt,Gies:2006cq}. In \cite{Schafer:2011as,Schafer:2015wta} the method was tested by computing the positive-energy conditions in various Casimir settings. These numerical methods are based on a Monte Carlo generation of worldline ensembles which, apart from providing an intuitive picture of the nonlocal nature of quantum fluctuations, is comparatively cheap due to its probabilistic nature (see \cite{Aehlig:2011xg,Mazur:2014gta,Edwards:2017won,Edwards:2019fjh}); we consider our analytic expressions could be used to test numerical computations in spherical geometries.

Most importantly, worldline numerics has been mostly applied to rigid boundaries but it has not been implemented under Robin conditions yet. As we mentioned, the technique presented in this article in combination with those in \cite{CMS,Bastianelli:2007jr,Vinas:2010ix} could provide a concrete realization for Robin boundary conditions in a worldline scheme. This motivates studies in this direction. Note finally that our method admits generalizations to other geometries, including curved space-times where the analysis of energy conditions is relevant. Research along these lines is currently in progress.

\section*{Acknowledgements}

LM acknowledges support from Consejo Interuniversitario Nacional under Programa de Becas EVC-CIN (Res.\ P.\ N$^\circ$ 403/18). Research of PP was partially supported by Universidad Nacional de La Plata under project 11/X615. IH acknowledges support from CONACYT through the SNI program.

\appendix

\section{Some correlation functions}\label{corre}

In this appendix we summarize some useful correlation functions,
\begin{align}
&\left\langle e^{-i\omega q_2(\tau)}\right\rangle
=\langle 1\rangle\,e^{-\frac{h(x)\,G}2\,\omega^2}\,,\label{exp}
\end{align}
\begin{align}
&\left\langle q_i(\tau)\,e^{-i\omega q_2(\tau)}\right\rangle
=-\langle 1\rangle\,\delta_{i2}\,i\omega\,h(x)\,G\,e^{-\frac{h(x)\,G}2\,\omega^2}
\,,\label{q-exp}
\end{align}
\begin{align}
&\left\langle p_i(\tau)\,e^{-i\omega q_2(\tau)}\right\rangle
=\langle 1\rangle\,\delta_{i2}\,\frac12\,\omega\,\mbox{}^{\bullet}G\,e^{-\frac{h(x)\,G}2\,\omega^2}\,,\label{p-exp}
\end{align}
\begin{align}
&\left\langle p_i(\tau)q_j(\tau)\,e^{-i\omega q_2(\tau)}\right\rangle=\langle 1\rangle\,\frac12\,i\,\mbox{}^{\bullet}G
\left(\delta_{ij}-\delta_{i2}\delta_{j2}\,\omega^2h(x)\,G\right)
e^{-\frac{h(x)\,G}2\,\omega^2}\,,\label{pq-exp}
\end{align}
\begin{align}
&\left\langle p^2(\tau)\,e^{-i\omega q_2(\tau)}\right\rangle
=\langle 1\rangle\,\left(\frac{1}{h(x)}+\frac14\,\omega^2\,\mbox{}^{\bullet}G^2\right)
e^{-\frac{h(x)\,G}2\,\omega^2}\,,\label{p2-exp}
\end{align}
\begin{align}
&\left\langle p^2(\tau)\,q_{i}(\tau)\,e^{-i\omega q_2(\tau)}\right\rangle
=-\langle 1\rangle\,\delta_{i2}\,i\omega
\left(G-\frac12\mbox{}^{\bullet}G^2+\frac14\,h(x)\omega^2\,G\ \mbox{}^{\bullet}G^2\right)
e^{-\frac{h(x)\,G}2\,\omega^2}\,,
\end{align}
\begin{align}
&\left\langle p^2(\tau)\,q_{i}(\tau)q_{j}(\tau)\,e^{-i\omega q_2(\tau)}\right\rangle
=\langle 1\rangle\left\{
\delta_{ij}\,\left(G-\frac12\,\mbox{}^{\bullet}G^2+\frac14\,h(x)\omega^2\,G\ \mbox{}^{\bullet}G^2\right)+\mbox{}\right.\nonumber\\[2mm]
&\left.\mbox{}-\delta_{i2}\delta_{j2}\,h(x)\omega^2\,G
\left(G-\mbox{}^{\bullet}G^2+\frac14\,h(x)\omega^2\,G\ \mbox{}^{\bullet}G^2\right)\right\}
e^{-\frac{h(x)\,G}2\,\omega^2}\,,
\end{align}
\begin{align}
&\left\langle p^2(\tau)\,q_i(\tau)q_j(\tau)q_k(\tau)\,e^{-i\omega q_2(\tau)}\right\rangle
=-\langle 1\rangle\,ih(x)\,\omega\,G\ e^{-\frac{h(x)\,G}2\,\omega^2}\times\mbox{}\nonumber\\[2mm]
&\mbox{}\times\left\{(\delta_{ij}\delta_{k2}+\delta_{ik}\delta_{j2}+\delta_{jk}\delta_{i2})
\left(G-\mbox{}^{\bullet}G^2+\frac14\,h(x)\omega^2\,G\ \mbox{}^{\bullet}G^2\right)+\mbox{}\right.\nonumber\\[2mm]
&\left.\mbox{}-\delta_{i2}\delta_{j2}\delta_{k2}\,h(x)\omega^2\,G
\left(G-\frac32\,\mbox{}^{\bullet}G^2+\frac14\,h(x)\omega^2\,G\ \mbox{}^{\bullet}G^2\right)\right\}\,.
\end{align}
In these expressions we use the notation $G=G(\tau,\tau)$ and $\mbox{}^{\bullet}G=\mbox{}^{\bullet}G(\tau,\tau)$.

\section{Expectation value of the delta-funcion}\label{B}

In this section we present an alternative calculation, technically more intuitive, of the leading contributions of the expectation values of the $\delta$-terms. Let us consider the direct contribution (given by \eqref{dirdelta}),
\begin{align}\label{alt}
\frac T2\int_{\mathbb{R}\times\mathbb{R}^+} dx_1dx_2
\ \left(1+x_1^2\right)\int_0^1d\tau\left\langle\delta\left(x_2+\sqrt T q_2(\tau)\right)\right\rangle\,.
\end{align}
The expectation value of the delta-function can be understood as the transition amplitude of a free particle from the point $x=(x_1,x_2)$ to some point at the boundary in dimensionless proper time $\tau$, and then back to the initial point $x$ in time $1-\tau$. We decompose this transition accordingly: we introduce an auxiliary delta-function $\delta(\sqrt T q_1(\tau)-\bar x_1)$ to enforce the particle to be at $(x_1+\bar x_1,0)$ in time $\tau$ and then integrate in the coordinate $\bar x_1$ along the boundary,
\begin{align}
\langle\delta(x_2+\sqrt T q_2(\tau))\rangle
&=\int_{-\infty}^\infty d\bar x_1
\ \frac{e^{-\frac{\bar x_1^2+x_2^2}{4Th(x)\tau}}}{4\pi T h(x)\tau}
\ \frac{e^{-\frac{\bar x_1^2+x_2^2}{4Th(x)(1-\tau)}}}{4\pi Th(x)(1-\tau)}\nonumber\\[2mm]
&=\frac{1}{\sqrt{\tau(1-\tau)}}\,\frac{e^{-\frac{x_2^2}{4Th(x)\tau(1-\tau)}}}{\sqrt{4\pi Th(x)}^3}\,.
\end{align}
Replacing this expectation value into \eqref{alt} one gets the correct contribution $\frac14$. The same procedure works for the contribution of the indirect $\delta$-terms. This would also work for the expectation values of $\theta$-terms were it not for the presence of the worldline field $p^2(\tau)$, for which it is not obvious how to generalize the present trick. However, we think it might be an interesting problem worth addressing.



\end{document}